\documentclass[prd,twocolumn,showpacs,showkeys,amsmath,floatfix,byrevtex]{revtex4}

\usepackage{graphicx}
\usepackage{dsfont}    % for `\realline' (see below)
\usepackage{dcolumn}   % for decimal point aligned tables

\providecommand{\realline}{\mathds{R}}
\providecommand{\earth}{{\scriptscriptstyle\oplus}}
\providecommand{\sunmass}{\mathrm{M}_{\scriptscriptstyle\odot}}
\providecommand{\prob}{\mathrm{P}}
\providecommand{\differential}{\mathrm{d}}
\providecommand{\likeli}{\mathcal{L}}
\providecommand{\degree}{$^\circ$}
\providecommand{\hour}{$^\mathrm{h}$}
\providecommand{\minute}{$^\prime$}
\providecommand{\differential}{\mathrm{d}}

\newenvironment{tinylist} 
   {\begin{list}          
      {-}{\setlength{\itemsep}{0.1\itemsep}
          \setlength{\topsep}{0.5\topsep}}}
   {\end{list}}

\begin{document}
  \title{Coherent Bayesian inference on compact binary inspirals\\
         using a network of interferometric gravitational wave detectors}
  \date{\today}
  \author{Christian R\"{o}ver}
  \author{Renate Meyer}
  \affiliation{Department of Statistics, The University of Auckland, Auckland, New Zealand}
  \author{Nelson Christensen}
  \affiliation{Physics and Astronomy, Carleton College, Northfield, MN, USA}

\begin{abstract}
Presented in this paper is the description of a Markov chain Monte Carlo (MCMC) 
routine for conducting 
coherent parameter estimation for interferometric gravitational wave observations 
of an inspiral of binary compact objects using %data from 
multiple detectors. 
%The MCMC technique uses data from several interferometers and infers 
%all nine of the parameters (ignoring spin) associated with the binary system, 
%including the distance to the source, the masses, and the location on the sky.
Data from several interferometers are processed, and all nine parameters (ignoring spin) 
associated with the binary system are inferred, including the distance to the source, 
the masses, and the location on the sky.
%The Me\-trop\-o\-lis-algorithm utilises advanced MCMC techniques, such as 
%importance resampling and parallel tempering. 
The data is 
%compared 
matched 
with time-domain inspiral templates that are 2.5 post-Newtonian (PN) in phase 
and 2.0~PN in amplitude. 
% new wording:
We designed and tuned an MCMC sampler so that it is able 
to efficiently find the posterior mode(s) in the parameter space 
and perform the stochastic integration
necessary for inference within a Bayesian framework.
Our routine could be implemented as part of an inspiral 
detection pipeline for a world wide network of detectors. 
Examples are given for simulated signals and data as seen by the LIGO and Virgo 
detectors operating at their design sensitivity.
\end{abstract}

% 02.70.Uu  Applications of Monte Carlo methods 
% 04.80.Nn  Gravitational wave detectors and experiments
% 05.45.Tp  Time series analysis 
% 07.05.Kf  Data analysis: algorithms and implementation; data management 
% 97.80.-d  Binary and multiple stars 


\maketitle

\section{Introduction}
The era of gravitational wave astronomy is now close upon us as numerous 
interferometric detectors are operating. The  Laser Interferometer Gravitational 
Wave Observatory (LIGO) \cite{Barish1997,BarishWeiss1999,AbbottEtAl2004a} 
has now reached its target sensitivity, and there is the hope that a detection
could come at any time \cite{KalogeraEtAl2006}. 
Around the globe a world-wide network of detectors is coming on-line; 
Virgo in Italy \cite{Brillet1997,CaronEtAl1996,AcerneseEtAl2006}, 
GEO in Germany \cite{HoughEtAl1997,LueckEtAl2006}, 
and TAMA in Japan \cite{TsubonoEtAl1997,TakahashiEtAl2003} 
are operating alongside LIGO in the quest for gravity wave detection. 
These ground based laser interferometers are sensitive 
to gravitational radiation in the frequency band from 40~Hz up to 8~kHz.

Coalescing binaries containing neutron stars or black holes promise to be one 
of the cleanest and most probable sources of detectable radiation \cite{Thorne1987}. 
Observation of inspiral events could provide important information on the structure 
of neutron stars \cite{CutlerEtAl1993,Hughes2002}. Even cosmological information can be 
extracted from the observation of inspiral events 
\cite{Schutz1986,ChernoffFinn1993,Markovic1993,CutlerFlanagan1994,Finn1996}.
The characteristics of radiation in the post-Newtonian regime will provide insight 
into highly non-linear general relativistic effects, such as the observation of 
the formation of a Kerr black hole as the binary system 
decays \cite{CutlerFlanagan1994,FlanaganHughes1998a,FlanaganHughes1998b}. 
%The LIGO Scientific Collaboration (LSC)
%has been actively searching for binary inspiral events. Using the data from 
%LIGO's S2~run, it was possible to set an upper limit on the neutron star coalescence 
%rate of less than 50 per year per Milky Way equivalent galaxy \cite{AbbottEtAl2005a}. 
The LIGO Scientific Collaboration (LSC)
has been actively searching for binary inspiral events 
\cite{AbbottEtAl2004b,AbbottEtAl2005a}, 
as well as conducting searches in coincidence with TAMA
\cite{AbbottEtAl2006b}.
Using the data from LIGO's S2~run, it was possible to set an upper limit 
on the neutron star coalescence rate of less than 50 per year 
per Milky Way equivalent galaxy \cite{AbbottEtAl2005a}. 
The LSC has also conducted searches for binary inspiral signals from primordial 
black holes (0.2--1.0~$\sunmass$) in the halo of our galaxy \cite{AbbottEtAl2005b}, 
plus more massive black hole systems 
where component masses are in the 3--20~$\sunmass$ range \cite{AbbottEtAl2006a}.

Bayesian inferential methods provide a means to use data from interferometric 
gravitational wave detectors in order to extract the parameters of a binary inspiral
event. Markov chain Monte Carlo (MCMC) methods are a powerful computation 
technique for parameter estimation within this framework;
they are especially useful in applications where the number of parameters 
is large \cite{MCMCinPractice}. Nice descriptions of the positive aspects of a 
Bayesian analysis of scientific and astrophysical data are provided 
in \cite{Gregory2001,Loredo1992,Finn1997}. In previous work we have developed MCMC 
routines for extracting five parameters associated with a binary inspiral event 
from data generated by a single interferometric detector 
\cite{ChristensenMeyer2001,ChristensenMeyerLibson2004,RoeverMeyerChristensen2006a}. 
Our MCMC code took data from a single interferometer, Fourier transformed it into 
the frequency domain, and then compared the result with 2.0~post-Newtonian (PN) 
stationary phase templates \cite{TanakaTagoshi2000}. 
One of the new methods that we implement in 
this current study, presented in this paper, is an MCMC routine that takes time 
domain interferometric data, and compares it to time domain templates that are 
2.5~PN in phase, and 2.0~PN in amplitude; a trivial modification of the code 
(though not implemented in the study presented here) 
%would be 
is
to extend the 
%complexity 
accuracy
of the signal waveforms to 3.5~PN in phase and 2.5~PN 
in amplitude 
\cite{BlanchetEtAl2002,BlanchetEtAl2005a,ArunEtAl2004,ArunEtAl2005,BlanchetEtAl2004}.
A critical task for a world-wide gravity wave detection network will be to not only 
detect a binary inspiral signal, but to say where it came from. 
For this purpose the LSC
has developed a \emph{coherent} binary inspiral search pipeline 
\cite{PaiDhurandharBose2001,PaiBoseDhurandhar2002,Bose2002}.
Coherent binary search pipelines and methods are also being developed within the
Virgo collaboration \cite{BirindelliEtAl2006} and TAMA \cite{MukhopadhyayEtAl2006}.
Along similar lines, we have developed a 
coherent MCMC parameter estimation routine, and in this present paper we describe it 
and provide results from a test on simulated data. 
The simulations involve binary neutron star 
inspirals observed by three well-separated interferometers: the 4~km LIGO detectors at Hanford, WA 
and Livingston, LA, plus the 3~km Virgo detector in Cascina, Pisa, Italy. 
The synthetic data for the LIGO and Virgo detectors 
has Gaussian noise with power spectral densities (PSD) 
that match their target sensitivities \cite{Sigg2004,AcerneseEtAl2005}. 
The MCMC code takes data from several interferometers, and estimates the two 
individual masses, time and phase at coalescence, distance to source, gravity wave 
polarization angle, angle of inclination of the binary system's orbital plane, 
and sky position in right ascension and declination. 
%The additional parameters (polarisation, inclination, position) 
%greatly complicate the algorithm, and create a vast parameter space. 
% ---
%The additional parameters polarisation, inclination, and sky location
%only become attainable when data from more than one interferometer are considered;
%they also greatly inflate the parameter space and with that complicate the analysis.
The additional parameters of polarization, inclination angle, and sky location
can only be estimated accurately when data from more than one interferometer
are considered; they also greatly inflate the parameter space and therefore
complicate the analysis.
MCMC methods have also been tested in a similar context
to recover the nine parameters of a binary black hole coalescence
%in LISA data \cite{Cornish2006}; 
%however, the problem setting is different
%(due to the different instrument, consequently a longer observation period, 
% lower frequencies being investigated, and a 2.0~PN phase model),
%and MCMC techniques are employed rather for
%optimisation than for integration, and on a subset of the
%parameters while others are solved for analytically.
in LISA data \cite{CornishPorter2007}.
However, the problem setting is different 
(due to the different instrument, longer observation period, 
lower frequencies being investigated, and the 2.0 PN phase model), 
and MCMC techniques were employed rather for optimization
than for integration; 
the MCMC was also only applied on a subset of the parameters,
while others were solved for analytically.

The organisation of this paper is as follows. 
After a brief introduction to the analysis problem %in Sec.~\ref{sec:AnStratModel}, 
we describe our approach alongside more detail 
on the applied model in Sec.~\ref{sec:AnStratModel}\@.
Sec.~\ref{sec:Implement} provides practical directions how we implemented MCMC methods 
in order to analyze data within the described framework.
Sec.~\ref{sec:Example} eventually illustrates results of applications
of our code to simulated data. 
We conclude the paper with a discussion and outlook in Sec.~\ref{sec:Discuss}.

\section{\label{sec:AnStratModel}Analysis strategy and modelling}
\subsection{Measuring gravitational waves}
In an inspiralling binary system, the two companions orbit around 
their centre of mass with decreasing orbital distance and period, 
until the system eventually collapses.
The gravitational radiation emitted by the system 
exhibits a `chirp' form, that is, 
an oscillation of increasing frequency and amplitude.
A laser interferometer is sensitive to space distortions along 
the directions of its two orthogonal arms, 
as it monitors the phase difference between the two laser beams 
that travel along the arms.
A gravitational wave is a quadrupole wave that is characterized 
by its direction of travel, polarization angle, 
and its two polarization amplitudes.
Its effect on a laser interferometer's measurements 
then is a linear combination of the effects
associated with the two polarizations, 
depending on the orientation of the interferometer with respect to the passing wave. 
Actual measurements, of course, are also exposed to noise.

%Measurements of a binary inspiral's chirp signal by a single interferometer
%will not be sufficient to infer all of the parameters that determined the
%signal's waveform and the interferometer's response. 
%Measurements from several seperate interferometers will 
%in general be required to derive e.g.\ the wave's direction of travel 
%by matching possible mutual orientations as well as 
%different arrival times of the signal at the different sites.
%Combining measurements from several interferometers will also enhance
%sensitivity and signal-to-noise ratio 
%\cite{PaiDhurandharBose2001}.
Measurements of a binary inspiral's chirp signal by a
single interferometer will not be sufficient to infer all of
the parameters that determine the signal's waveform
and the interferometer's response. Measurements from
several separate interferometers will, in general, be required
to derive (for example) the wave's direction of travel by
matching possible mutual orientations as well as different
arrival times of the signals at the different sites. Combining
measurements from several interferometers will also
enhance sensitivity and signal-to-noise ratio
\cite{PaiDhurandharBose2001}.

\subsection{The Bayesian approach}
We apply a \emph{Bayesian} approach to this inferential problem, 
that is, the term `\emph{probability}' is used in a broader sense than 
in the more common `\emph{frequentist}' interpretation 
\cite{Jaynes1986,Jaynes,Finn1997}. 
Probability calculus here is applied to process and infer 
\emph{states of incomplete information} that are reflected by probability distributions,
and that are conditional on prior knowledge and/or the data at hand.
This allows one to treat unknown parameters as random variables 
that follow a \emph{prior distribution} representing 
the researcher's pre-experimental knowledge and uncertainty.
The gain in information through observation of data 
then follows in a straight-forward fashion through
the application of Bayes' theorem,
yielding the \emph{posterior distribution} of the parameters.
The posterior distribution, which is essentially the product
of the parameters' prior distribution and the likelihood of the data,
then poses the basis for inference \cite{BDA}.

Inference through the posterior distribution usually involves the 
solution of integrals, since one is typically interested in figures 
such as the marginal (posterior) expectations of individual parameters,
marginal (posterior) densities,
or (posterior) probabilities of certain events, which are derived
from the posterior distribution by integrating over the parameter 
space.
In many cases when analytic integration is not possible, numerical 
methods are employed, usually (pseudo\mbox{-)} stochastic techniques
like Markov chain Monte Carlo (MCMC) methods that simulate random
draws from the posterior ditribution, then allowing one to approximate
the desired integrals by sample statistics
\cite{BDA,MCMCinPractice}.

\subsection{Parameters} \label{sec:parameters}
The waveform that is \emph{measured} at a certain interferometer depends 
on the characteristics of the inspiral event as well as the  
orientation of the source relative to the interferometer. 
The nine `global' parameters determining the response 
of Earth-bound interferometers are:
\begin{tinylist}
  \item individual masses ($m_1, m_2 \in \realline^+;\; m_1\leq m_2$),
  \item luminosity distance ($d_L \;\in\; \realline^+$),
  \item inclination angle ($\iota \;\in\; [0,\pi]$),
  \item coalescence phase ($\phi_0 \;\in\; [0,2\pi]$),
  \item coalescence time at geocenter ($t_c \;\in\; \realline$),
  \item declination ($\delta \;\in\; [-\frac{\pi}{2},\frac{\pi}{2}]$),
  \item right ascension ($\alpha \;\in\; [0,2\pi]$) and
  \item polarization ($\psi \;\in\; [0,\pi]$),
\end{tinylist}
the latter four of which
affect the measurement at the \mbox{$I$-th} detector 
in terms of the `local' parameters
\begin{tinylist}
  \item local coalescence time ($t_c^{(I)} \;\in\; \realline$),
  \item altitude ($\vartheta^{(I)} \;\in\; [0,\pi]$),
  \item azimuth ($\varphi^{(I)} \;\in\; [0,2\pi]$) and
  \item polarization ($\psi^{(I)} \;\in\; [0,\pi]$).
\end{tinylist}
These `local' parameters are derived from the %mutual
lo\-ca\-tions/o\-ri\-en\-ta\-tions of the source and the individual interferometers
with respect to each other.
For more specific definitions and conventions see e.g.~\cite{Blanchet2001}.
In the following we will refer to the two parameter sets as the 
\emph{global} parameter vector 
\begin{subequations}
\begin{equation} \label{eqn:globalpar}
  \theta^\earth = (m_1, m_2, d_L, \iota, \phi_0, t_c, \delta, \alpha, \psi),
\end{equation}
and the \emph{local} parameter vector
\begin{equation} \label{eqn:localpar}
  \theta^{(I)} = (m_1, m_2, d_L, \iota, \phi_0, t_c^{(I)}, \vartheta^{(I)}, \varphi^{(I)}, \psi^{(I)})
\end{equation}
\end{subequations}
with respect to a specific interferometer~$I$.
Not all of the above parameters will usually be of primary interest; 
especially coalescence phase $\phi_0$, polarization $\psi$ 
or inclination $\iota$ might be regarded as nuisance parameters.

\subsection{Network likelihood}
An interferometer's data output~$z$ 
is assumed to be the sum of the actual signal~$s(\theta)$, 
depending on the true parameters~$\theta$,
and (interferometer-specific) coloured noise.
The (real-valued) data~$z$ and signal waveform~$s(\theta)$ 
enter the likelihood function in terms of 
their (complex-valued) Fourier transforms~$\tilde{z}$ and $\tilde{s}(\theta)$,
the noise is specified through its power spectral density (PSD) $S_n$.
%The likelihood function for a specific interferometer~$I$ 
%is (up to a normalising constant) proportional 
%to the sum of squared and normalised differences between
%the Fourier transforms of observed signal~($\tilde{z}$)
%and signal template ($\tilde{s}(\theta)$) 
%over the discrete set of Fourier frequencies 
%$\{(i\times\Delta_f):\;i_L\leq i\leq i_U\}$:
The likelihood function for a specific interferometer~$I$ 
then is (up to a normalising constant) proportional 
to the following expression
\begin{equation} \label{eqn:detectorlikelihoodInt}
  \likeli^{(I)}\big(\theta^{(I)}\big) \propto
  \exp\left( -2 \int_0^\infty
  \!\frac{|\tilde{z}(f)-
         \tilde{s}(f,\theta^{(I)})|^2}
       {S_n(f)} \differential f
  \right)
\end{equation}
\cite{FinnChernoff1993}.
For actual data, discretized and measured 
over a finite interval of length~$\delta_t$,
it is computed as
the sum of squared and normalized differences between
the Fourier transforms of the observed signal~($\tilde{z}$)
and the signal template ($\tilde{s}(\theta)$) 
over the discrete set of Fourier frequencies 
$\{(i\times\Delta_f):\;i_L\leq i\leq i_U\}$:
\begin{equation} \label{eqn:detectorlikelihoodSum}
  \likeli^{(I)}\big(\theta^{(I)}\big) \propto
  \exp\Biggl(\frac{-\!2}{\delta_t}\sum_{i=i_{L}}^{i_{U}}
  \!\frac{|\overbrace{\tilde{z}(i\!\times\!\Delta_f)}^{\text{data}}-
         \overbrace{\tilde{s}(i\!\times\!\Delta_f,\theta^{(I)})}^{\text{template}}|^2}
       {\underbrace{S_n(i\!\times\!\Delta_f)}_{\text{noise PSD}}}
  \Biggr)
\end{equation}
where $i_L\times\Delta_f$ and $i_U\times\Delta_f$ are the lower 
and upper bounds of the examined frequency range. %,
%and $\delta_t$ is the length of the analysed data segment 
Note that, although not labeled as such here, data~$z$, 
noise spectrum~$S_n$, etc.\
are specific for the different interferometers~$I$.
Assuming that noise is independent across different interferometers, 
the network likelihood then is the product of the individual
interferometer likelihoods:
\begin{equation} \label{eqn:networklikelihood}
  \likeli(\theta^\earth) \;=\; \prod_I \likeli^{(I)}\big(\theta^{(I)}\big).\\
\end{equation}

\subsection{Prior specification}
%The prior information is straightforwardly specified for the `geometrical'
The prior information is specified in a straightforward fashion 
for the `geometrical' parameters that determine the location 
and orientation of the inspiral event.
A~priori, the event is assumed to be equally likely 
across all possible directions; this leads to a uniform prior for the 
right ascension~$\alpha$, and a prior density
\begin{equation} \label{eqn:alphaprior}
  \textstyle
  f(\delta) =  \frac{1}{2} \cos(\delta) 
\end{equation}
that is proportional to the circumference of
the corresponding `circle of latitude' 
for the declination~$\delta$.
Analogously, the prior density of the inclination~$\iota$ is 
\begin{equation} \label{eqn:iotaprior}
  \textstyle
  f(\iota) =  \frac{1}{2} \sin(\iota),
\end{equation}
and the remaining parameters, polarization~$\psi$ and 
coalescence phase~$\phi_0$, have uniform priors.
The prior specifications for these parameters may also be regarded
as \emph{Maximum Entropy} choices 
\cite{Jaynes1968,Jaynes}.
% InfoTheory, p.269, example 11.2.4
% uniform=MaxEnt on unit sphere: see Mardia1975, p.352, or MardiaJupp, p.172.
\\

The coalescence time~$t_c$ is assumed to be known in advance up to a certain 
accuracy through preprocessing of the data 
\cite{AbbottEtAl2004b,MarionEtAl2003,AmicoEtAl2003};
for now we set its prior to be uniform across 
$\pm 5 \mbox{ms}$ around the true value, 
which of course is known for simulated data.

The joint prior of the remaining parameters, masses $m_1$, $m_2$ 
and luminosity distance $d_L$, is set in order to reflect the
distribution of parameters \emph{given} the event has been detected 
in the first place.
Initially, the prior for the two inspiral companions' individual masses 
is assumed to be uniform between 0.6 and 3.0~$\sunmass$ 
(solar masses: $\sunmass\approx 2 \!\times\!10^{30}\;\mbox{kg}$), 
which effectively covers the range of values expected for neutron stars.
The prior for $d_L$ is derived from the assumption that inspirals happen 
uniformly across space, so that $\prob(d_L\leq x) \propto x^3$. 
So far, this leads to an improper distribution (that has an infinite integral).
But inspiral signals obviously cannot originate from arbitrarily 
great distances, since at some point their signals become too faint to be detected.
We incorporated this restriction by taking into consideration 
the \emph{detection probability}~$D$, which we assume to depend on the signal's
\emph{amplitude}, which is roughly proportional to
\begin{equation} \label{eqn:amplitude}
  \mathcal{A}(m_1, m_2, d_L)
    =  \ln\Biggl(\frac{\sqrt{m_1 m_2}}{d_L\,(m_1+m_2)^{\frac{1}{6}}}\Biggr),
\end{equation}
neglecting for simplicity the effects of orientation parameters
($\mathcal{A}$~is actually the logarithmic amplitude) 
\cite{RoeverMeyerChristensen2006a}.
We could have set a threshold amplitude below which inspirals would be assumed
undetectable, but favoured a `smoother' transition that does not 
explicitly apply zero probability to parts of the parameter space. 
Instead we model the dependence between signal amplitude and detection probability 
using a (sigmoidal) logistic function of the form
\begin{equation} \label{eqn:logisticfun}
  D_{a,b}(x) = \frac{1}{1+\exp(\frac{x-a}{b})}
\end{equation}
whose parameters $a$ and $b$ are set so that 
$D_{a,b}(x_U)=1-p$ and $D_{a,b}(x_L)=p$, for some upper and lower reference points
$x_U$ and $x_L$, and some $0<p<0.5$ (e.g.\ $p:=0.1$).
So $x_U$ denotes the amplitude at which the detection probability reaches~$1-p$, 
and $x_L$ is the amplitude at which the probability falls below~$p$.
In order to fit $D$ through these points, its parameters are set to
\begin{equation} \label{eqn:logisticpar}
  a:=\frac{x_L+x_U}{2} \qquad \text{and} \qquad b:=\frac{x_U-x_L}{2\ln(\frac{p}{1-p})}.
\end{equation}
So the density of the resulting (proper) prior distribution 
of individual masses and distance is
\begin{equation} \label{eqn:logistic}
  f(m_1,m_2,d_L) \;=\; c \times d_L^2 \times 
                   D_{a,b}\bigl(\mathcal{A}(m_1,m_2,d_L)\bigr)
\end{equation}
for some normalizing constant $c\in\realline^+$
\cite{RoeverMeyerChristensen2006a}.
For the examples shown here, we set
$x_U:=\mathcal{A}(2.0 \sunmass, 2.0 \sunmass, 45 \mbox{Mpc})$,
$x_L:=\mathcal{A}(2.0 \sunmass, 2.0 \sunmass, 50 \mbox{Mpc})$ and
$p:=0.1$,  
so an optimally oriented 2.0-2.0~$\sunmass$ inspiral is assumed to be detectable 
out to~45 and 50~Mpc with probabilities of 90\% and 10\%, respectively.
\\

As a `side effect' of this prior definition, 
larger masses have a greater prior probability,
since inspirals involving large masses may originate from farther distances 
while low-mass inspirals need to be close in order to be observable at all. 
This feature is also known as the \emph{Malmquist effect};
incorporating it into the prior will compensate for selection bias 
that would otherwise affect the results \cite{Teerikorpi1997,Sandage2001}.
The definition of priors, especially for coalescence time~$t_c$, 
individual masses $m_1$ and $m_2$ and their relation to the luminosity distance~$d_L$,
and possibly also for the sky location $(\delta, \alpha)$,
may be refined at a later stage when e.g.\ some substantiated knowledge is available
about the performance of the upstream detection pipeline, 
which might provide rough estimates of some of the parameters 
together with the detection 
\cite{AbbottEtAl2004b,MarionEtAl2003,AmicoEtAl2003}.
For now we aim for simple and general formulations.

\section{\label{sec:Implement}Implementation}
\subsection{General}
In order to analyze data in terms of the above framework, 
we implemented an MCMC sampler in~C that is supposed to 
both \emph{find} the global mode(s) of the posterior distribution 
and then `\emph{explore}' the distribution,
i.e.\ simulate random draws from the posterior.
Data is imported from the \emph{Frame format} 
using the Frame Library \cite{FrameLibrary}.
Prior to the analysis, the data is filtered and downsampled
by a factor of~4
\cite{IEEE-DSP-8.2,ParksMcClellan}.
(Pseudo\mbox{-)} random number generation within the MCMC sampler
was implemented using Randlib \cite{randlib}.

The MCMC sampler writes only every 25th of the drawn samples to a text file, 
in order to reduce the effects of subsequently correlated samples,
and also to keep the data volume at a reasonable level.
The MCMC output then is imported into~R, a statistical software, for eventual analysis
\cite{R-Manual}.

The marginal densities that are shown in this paper are 
\emph{kernel density estimates} \cite{Scott}. 
% Epanechnikov kernel, Silverman bandwidth.
Two-dimensional densities are illustrated by greyscale plots 
(with darker areas corresponding to greater densities), 
%and in addition the contour line enclosing 95\%
%of the probability mass is shown.
and in addition the contour line enclosing the most probable region
(accumulating 95\% probability) is shown.

\subsection{Likelihood implementation}
In order to compute the coherent network-likelihood, 
first the individual interferometer-likelihoods need to be determined.
The primary `ingredients' for the in\-ter\-fe\-rom\-e\-ter-like\-li\-hood are
\begin{tinylist}
  \item the Fourier-transform of the data $\tilde{z}$,
  \item the noise spectrum $S_n$,
  \item the `local' parameter set $\theta^{(I)}$ and
  \item the (Fourier-transformed) signal template $\tilde{s}(\theta^{(I)})$,
\end{tinylist}
the first two of which only need to be determined once at the very beginning 
of the analysis, while the latter two (in general) need to be re-computed for each
likelihood evaluation.

For all (discrete) Fourier-transformations we use the freely available 
FFTW library \cite{FFTW}. The noise spectrum is estimated from a section of 
data that is disjoint from the actually analyzed data set \cite{Welch1967}.
In order to minimize undesirable leakage effects, the data is `windowed'
before Fourier-transformation; using a Hann window for spectrum estimation,
and a Tukey window for data and template transformations
\cite{Harris1978}.

Internally, along with the noise spectra, data Fourier transforms etc.\ 
corresponding to each of the interferometers~$I$, 
a set of vectors defining the interferometer's location and orientation is stored.
This allows to derive the interferometer-specific parameters 
(local coalescence time~$t_c^{(I)}$, altitude~$\vartheta^{(I)}$, azimuth~$\varphi^{(I)}$ 
and polarization~$\psi^{(I)}$) 
with respect to the galactic and Earth coordinate systems 
from the global parameter vector~$\theta^\earth$
via vector operations like rotations, orthogonal projections, etc.\ \cite{Allen1996,Lang1999,WGS84}.

\subsection{Time-domain waveform generation}
Template waveforms~$\tilde{s}(\theta)$ are generated in the time-domain
and then (numerically) Fourier transformed to the frequency domain.
Here we used waveform approximations 
%that are 2.0~PN in amplitude and 2.5~PN in phase.
%%%% swapped for consistency: %%%% 
that are 2.5~PN in phase and 2.0~PN in amplitude. 
% added:
The rather complex expressions for these are omitted here 
and can be found in \cite{Blanchet2001}.
We preferred working in the time-domain, since frequency-domain templates
might introduce discrepancies because they are exact analytic Fourier transforms 
of the `parametric' waveforms, 
while the actual data is of finite extent and affected by leakage 
introduced through the numerical discrete Fourier transformation.
When using time-domain templates and Fourier transfoming these, 
the resulting frequency-domain templates are the more accurate match 
to the Fourier-transformed data.
Another advantage of generating templates in the time-domain is 
the availability of a wider range of signal waveforms; 
the extension to higher PN appoximations 
(e.g.\ 3.5~PN phase and 2.5~PN amplitude
\cite{BlanchetEtAl2002,BlanchetEtAl2005a,ArunEtAl2004,ArunEtAl2005,BlanchetEtAl2004})
or the consideration of additional parameters 
(e.g.\ spin effects \cite{BuonannoChenVallisneri2003}) 
would be straightforward to implement.

\subsection{MCMC implementation}
In order to enhance the MCMC sampler's performance we applied reparametrisations
to some of the parameters. The individual masses ($m_1$, $m_2$) are highly correlated
in their posterior distribution \cite{CutlerFlanagan1994}, 
making sampling from the original parameters extremely difficult. 
%Re-expressing the masses in terms of \emph{chirp mass}
%($m_c=\frac{(m_1 m_2)^{3/5}}{(m_1+m_2)^{1/5}}$)
%and the (symmetric) \emph{mass ratio} ($\eta=\frac{m_1 m_2}{(m_1+m_2)^2}$)
%yields a posterior that is much easier to sample from.
Re-expressing the masses in terms of \emph{chirp mass}~$m_c$
and the (symmetric) \emph{mass ratio}~$\eta$, where
\begin{equation} \label{eqn:reparam}
  m_c=\frac{(m_1 m_2)^{3/5}}{(m_1+m_2)^{1/5}} 
  \quad \text{and} \quad 
  \eta=\frac{m_1 m_2}{(m_1+m_2)^2},
\end{equation}
yields a posterior that is much easier to sample from.
We then reparameterized the luminosity distance from $d_L$ to $\ln(d_L)$,
which implicitly yields an unbounded parameter space and proposal step widths that are
proportional to the current distance~$d_L$.
Declination~$\delta$ and inclination~$\iota$ were transformed to
$\sin(\delta)$ and $\cos(\iota)$, which leads to uniform prior distributions
over the new parameters.
\\

The MCMC algorithm was implemented as a Me\-trop\-o\-lis-sampler
\cite{BDA,MCMCinPractice}
that was extended to a parallel tempering algorithm.
The idea of tempering is to consider a smoothed (`tempered') version of the 
actual objective function (here the posterior distribution), or, analogously,
do its exploration (optimization, MCMC sampling,...) following `relaxed' rules.
In optimization contexts, the tempering is often faded out over time, 
the result being a `simulated annealing' algorithm, 
which starts off at a high temperature in order to find the global mode 
amongst other minor modes, but eventually ends up optimizing 
the original objective function.
%
%Parallel tempering is a special case of the
%`Me\-trop\-o\-lis-coupled MCMC' (MCMCMC) algorithm,
%in which several `tempered' chains are run in parallel, 
%and additional steps
%are introduced to allow for `swapping' between chains \cite{MCMCinPractice}.
Parallel tempering is a special case of the
`Me\-trop\-o\-lis-coupled MCMC' (MCMCMC) algorithm,
in which several `tempered' chains are run in parallel, 
each having a different (constant) temperature. 
So the tempering does not vary \emph{over time}, 
but instead is realised \emph{across parallel chains}, 
with additional proposals allowing for swapping between chains.
%and additional steps
%are introduced to allow for `swapping' between chains \cite{MCMCinPractice}.
Instead of sampling from the regular posterior distribution
with density function~$f$
(which is essentially the product of prior~$\pi$ and likelihood~$\likeli$: 
 $f(\theta)\propto\pi(\theta) \likeli(\theta)$),
the tempered chains sample from a modified distribution
\begin{equation}
  f_T(\theta) \,\propto\, \pi(\theta) \,\likeli(\theta)^\frac{1}{T},
\end{equation}
where $T\geq 1$ denotes the `temperature', and for which in the extreme cases
$T=1$ and $T\rightarrow\infty$ the tempered distribution~$f_T$ equals posterior and 
prior respectively. 
Chains running at higher temperatures can be considered as sampling from a 
`relaxed' or `widened' posterior which is then used 
as proposal distribution (through the swapping between chains) for `cooler' chains, 
thus improving both convergence and mixing.
The draws from the `coolest' chain with $T=1$ are the only ones that
are eventually used for posterior inference.
\\

The starting values for the sampler are determined using 
\emph{importance resampling} 
\cite{BDA}.  %,SmithGelfand1992}. 
% BDA, section 10.5, page 312 sqq.
In a simplified setting of the problem (considering only 5~parameters and 
one interferometer \cite{RoeverMeyerChristensen2006a}), 
this was sufficient to yield reasonable posterior samples 
that were close enough to the main mode so the sampler then converged reliably and fast.
Due to the much larger parameter space and the computationally 
more expensive likelihood, ensuring convergence through good starting values
is not feasible any more. 
Instead, convergence is now supported through the use of parallel tempering, 
which enables the sampler to cross gaps between (local) modes 
and eventually find the posterior's global mode(s).

As proposal distribution for the MCMC sampler we used 
a multivariate Student's $t$\mbox{-}distribution with 3~degrees of freedom.
In addition to these `regular' proposals, sometimes draws from the prior
are proposed for some parameters in order to enhance convergence, 
or steps to `related' parts of the parameter space, like a step 
from phase~$\phi_0$ to $\phi_0\pm\pi$, which corresponds to an (almost)
equally likely parameter combination if the two masses are (almost) equal.
Proposals like these are valid as long as a certain symmetry 
is maintained, i.e.\ every proposed step is as likely as the reverse step;
otherwise one would need to switch to a Metropolis-Hastings sampler 
that is able to account for asymmetric steps \cite{BDA,MCMCinPractice}.

\subsection{Signal-to-noise ratios}
The signal-to-noise ratios (SNR) stated in subsequent sections are 
defined as follows.
The interferometer-specific SNR
of a certain signal~$s(\theta^\earth)$ 
received at interferometer~$I$ and
embedded in noise
with spectral density~$S_n$ is defined as:
\begin{equation}
  \varrho^{(I)} 
     = 2\sqrt{\int_0^\infty\frac{|\tilde{s}(f,\theta^{(I)})|^2}{S_n(f)}\differential f}
\end{equation}
\cite{ChristensenMeyerLibson2004}. % p. 323
We computed it---in analogy to Eq.~(\ref{eqn:detectorlikelihoodSum})---over 
the same frequency range that is relevant for the likelihood.
The network SNR then is defined as:
\begin{equation}
  \varrho_{\text{network}}
     = \sqrt{\sum_I \left(\varrho^{(I)}\right)^2}
\end{equation}
\cite{CutlerFlanagan1994}.

\section{\label{sec:Example}Example application using simulated data}
\subsection{Overview}
In the following two sections we will illustrate results 
generated by our MCMC implementation on simulated data.
%Firstly, in section~\ref{sec:example2} the evidence on parameters 
Firstly, in section~IV.B the evidence on parameters 
in the measured data is illustrated for one simulated signal
by looking at the results of an MCMC run in detail.
%In the following section~\ref{sec:example3} the effect 
In the following section~IV.C the effect 
of varying signal properties on the evidence in the data 
is demonstrated and some peculiarities are pointed out.
The posterior distributions of parameters are compared across simulated signals 
that are observed at different distances (and with that, SNRs), 
but which are otherwise identical.
While the \emph{individual} SNRs (at each interferometer) 
for each of these examples are similar,
%differ about a factor of~2, 
a more extreme example with an almost zero SNR 
at one of the interferometers is considered as well.

\subsection{Recovering the inspiral's parameters}
\label{sec:example2}We simulated an inspiral event that is measured at 
three interferometer sites, namely Hanford (LHO), Livingston (LLO) and Pisa (V).
\begin{figure*}[t]
  \includegraphics[width=17cm]{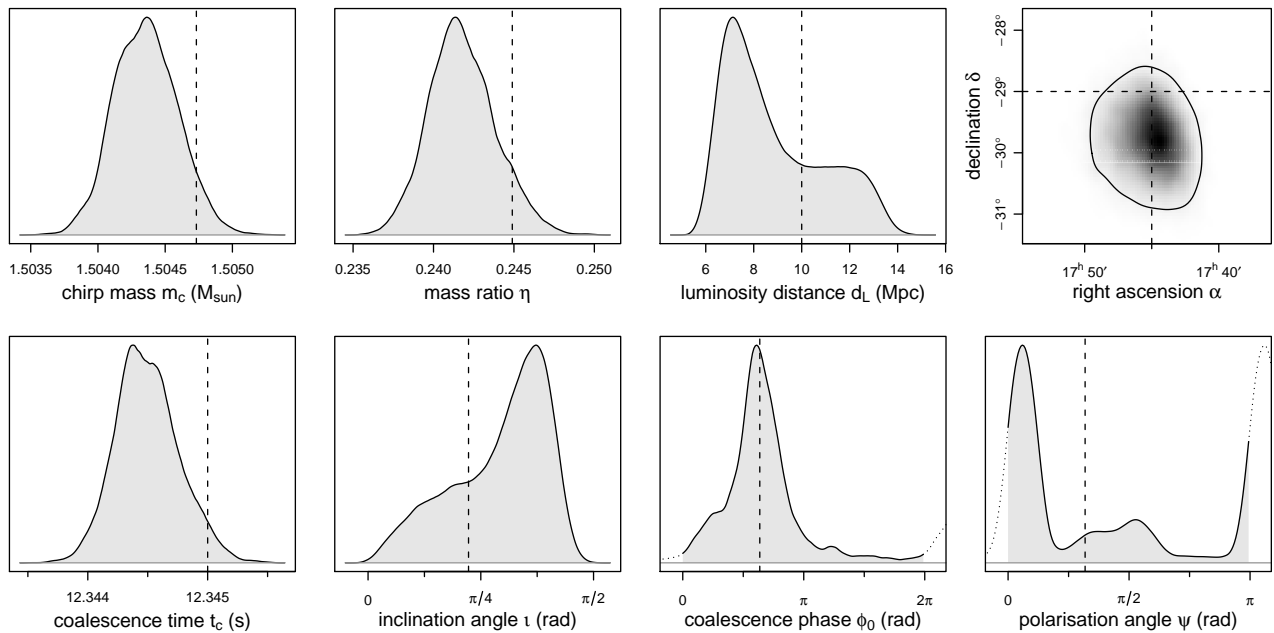}
  \caption{\label{fig:posteriordensities}Marginal posterior densities 
           of the inspiral's parameters for our example problem. 
           Dashed lines indicate the true values.}
\end{figure*}
Due to their different noise characteristics, the frequency ranges of the likelihoods
were set to 30--1600~Hz for the Virgo observatory, 
and 40--1600~Hz for the other two LIGO interferometers
(cp.\ Eq.~(\ref{eqn:detectorlikelihoodSum})).
The amount of data to consider was set to be the 40~seconds before coalescence for Virgo,
and 20~s for the other two. 
This matches the time an inspiral of this kind spends emitting radiation 
within the above frequency ranges, 
and would in a realistic search need to be set 
either with respect to worst-case considerations, 
or based on prior information supplied by the detection pipeline.
The original sampling rates of the data were $20\,000$~Hz (V) 
and $16\,384$~Hz (LHO, LLO), 
and the signals were superimposed with Gaussian noise matching the corresponding
interferometer's design sensitivities \cite{Sigg2004,AcerneseEtAl2005}.
\begin{table}[b]
  \caption{\label{tab:summarystats}
           Some key figures summarizing the marginal posterior distributions
           of individual parameters, where meaningful.
           Mean and median characterize the distributions' centers.
           Given the observed (simulated) data, the parameters fall 
           within the central posterior intervals 
           with 95\% probability. 
           The true parameter values used to generate the data 
           are shown as well.}
  \begin{ruledtabular}
  \begin{tabular}{lccccl}
             &  mean   &  median &     95\% c.p.i.    &  true   & unit \\ \hline
    $m_c$    &  1.5044 &  1.5044 &  [1.5039,  1.5048] &  1.5047 & $\sunmass$ \\
    $\eta$   &  0.2418 &  0.2417 &  [0.2380,  0.2460] &  0.2449 &            \\ 
    $t_c$    & 12.3445 & 12.3445 & [12.3440, 12.3450] & 12.3450 & s          \\
    $d_L$    &  8.89   &  8.29   &    [6.25, 13.1]    & 10.00   & Mpc        \\
    $\delta$ & -29.76\degree & -29.77\degree 
              & [-30.74\degree, -28.84\degree] & -29.00\degree & \\
    $\alpha$ & 17\hour 45.0\minute & 17\hour 44.9\minute 
              & [17\hour 42.1\minute, 17\hour 48.9\minute] & 17\hour 45.0\minute & \\
    $\iota$  &  0.911  &  0.995  &  [0.194, 1.354]    & 0.700   & rad \\[1ex]
    $m_1$    &  1.446  &  1.442  &  [1.389, 1.526]    & 1.5     & $\sunmass$ \\
    $m_2$    &  2.080  &  2.084  &  [1.965, 2.170]    & 2.0     & $\sunmass$ \\
    $m_t$    &  3.526  &  3.527  &  [3.490, 3.559]    & 3.5     & $\sunmass$ \\
  \end{tabular}
  \end{ruledtabular}
\end{table}
The example inspiral had parameter values of
$d_L=10\mbox{~Mpc}$ for the distance
and  masses of $m_1=1.5$~$\sunmass$ and $m_2=2.0$~$\sunmass$
(chirp mass~$m_c=1.505$~$\sunmass$, mass ratio~$\eta=0.245$).
The resulting SNRs at the three interferometer sites
are: LHO:~16.4, LLO:~21.2 and V:~12.6 (network SNR:~29.6).

Six parallel MCMC chains were run within the parallel tempering scheme.
With this amount of data the MCMC code generated
some 80 samples per minute on a 3.2~GHz Pentium desktop PC, 
so considering the parallel chains (six) and the thinned output (every 25th), 
an actual posterior sample is generated every 113~seconds.
The first chain of the parallel tempering MCMC sampler 
converged after some $75\,000$~iterations, 
after `thinning out' of the samples and discarding the burn-in phase, 
the resulting posterior sample size was $12\,500$~samples.

FIG~\ref{fig:posteriordensities} shows marginal posterior densities 
of the nine parameters %, 
for our example problem,
and Table~\ref{tab:summarystats} lists some numerical posterior estimates.
\begin{figure}[h]
  \includegraphics[width=8cm]{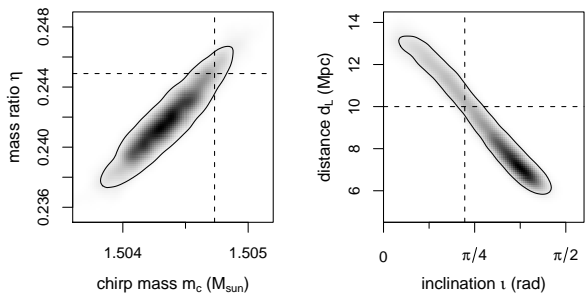}
  \caption{\label{fig:bivariatePosteriors}Marginal joint posterior distributions
           of two pairs of parameters. Dashed lines indicate the true values.}
\end{figure}
Although correlations between parameters are already greatly reduced
through the reparametrisation, some correlation still remains.
FIG.~\ref{fig:bivariatePosteriors} illustrates correlations between two 
such pairs of parameters, %:
one can see that the `new' mass parameters $m_c$  and~$\eta$ are still dependent
(though orders of magnitude less than $m_1$ and $m_2$ were), 
and also that the uncertainty in the luminosity distance~$d_L$ 
is tied to the uncertainty in the inclination angle~$\iota$, 
%since these two parameters can compensate or mimic each other to a certain degree.
since these two parameters can compensate or mimic each other's effect to a certain degree.

Distributions of other variables derived from the parameters could
be investigated, their distributions then depending on the \emph{joint} distribution 
of the involved parameters \cite{RoeverMeyerChristensen2006a}.
Examples would be the individual masses ($m_1$, $m_2$), or the total mass~$m_t=m_1+m_2$;
the distribution of $m_t$ is narrower than those of both $m_1$ and $m_2$, 
due to their negative correlation (cp.\ Table~\ref{tab:summarystats}).

\subsection{Results for varying signal characteristics}
\label{sec:example3}We performed additional runs with varying settings 
of the `true' parameters of the simulated signal.
As one would expect, the precision of parameter estimation is proportional 
to the signal's strength; Table~\ref{tab:precisions} shows 
the standard deviations of some of the parameters' posterior distributions.
The posterior is narrowest for a close-by inspiral
of high masses, and gets wider for both lower mass or greater distance. 

These results are in agreement with earlier estimates 
of the accuracy to be expected from such parameter estimates \cite{CutlerFlanagan1994}. 
The great difference in relative accuracies of parameters related to phase evolution 
(like chirp mass~$m_c$ and reduced mass~$\mu=\frac{m_1m_2}{m_1+m_2}=m_t \eta$)
versus those affecting the signal's amplitude (like distance~$d_L$) is confirmed,
%as well as the correlation between $m_c$ and~$\mu$.
and the correlation between $m_c$ and~$\mu$ is verified as well.

\begin{figure}[b]
  \includegraphics[width=8cm]{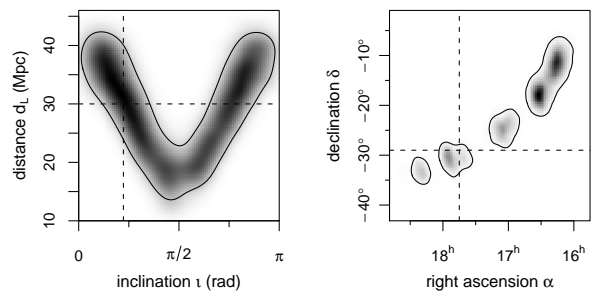}
  \caption{\label{fig:ambiguities}At greater distance the `orientation' of
           the inclination angle~$\iota$ cannot be resolved any more,
           both directions are roughly equally likely 
           ($\prob(\iota<\frac{\pi}{2}) \approx \frac{1}{2} \approx \prob(\iota>\frac{\pi}{2})$).
           At the same time, with the lower SNR the sky location's posterior turns multimodal.
           (Dashed lines indicate the true values.)}
\end{figure}

At decreasing SNRs, certain parameters cannot be determined unambiguously
any more. 
\begin{table*}
  \caption{\label{tab:precisions}Individual and total SNRs for different signals,
           and some characteristics of the resulting posterior distributions.
           The accuracy of some of the parameters is illustrated by the posterior standard deviations
           for $(\delta,\alpha)$, $t_c$, $d_L$, $m_c$ and~$\mu$ (percentages refer to the true value).
           The correlation coefficient for $m_c$ and $\mu$ 
           shows the (posterior) interdependence between the two parameters. 
           Our results are consistent with those presented in \cite{CutlerFlanagan1994}.} 
  \begin{ruledtabular}
    \begin{tabular}{cccddddcccccccc}
                  %   I    I     I
       masses             & distance     & & \multicolumn{4}{c}{signal-to-noise ratios}& &\multicolumn{5}{c}{posterior standard deviations}& & \\
       \cline{4-7} \cline{9-13}
       $m_1$-$m_2$        & $d_L$        & & \multicolumn{1}{c}{LHO} 
                                           & \multicolumn{1}{c}{LLO} 
                                           & \multicolumn{1}{c}{V}
                                           & \multicolumn{1}{c}{network}
                                         & & $(\delta,\alpha)^\mathrm{a}$ & $t_c$ & $d_L$ & $m_c$ & $\mu$
                                         & & $\mathrm{Cor}(m_c,\mu)$ \\ \hline
       1.5-2.0$\;\sunmass$ & $10\;$Mpc & & 16.4 & 21.2 & 12.6 & 29.6 & & $0.011\;$rad & $0.26\;$ms & $20\;$\% & $0.016\;$\% & $0.35\;$\% & & 0.95\\
       1.5-2.0$\;\sunmass$ & $20\;$Mpc & &  8.2 & 10.6 &  6.3 & 14.8 & & $0.030\;$rad & $0.49\;$ms & $25\;$\% & $0.031\;$\% & $0.69\;$\% & & 0.94\\
       1.5-2.0$\;\sunmass$ & $30\;$Mpc & &  5.5 &  7.1 &  4.2 &  9.9 & & $0.207\;$rad & $1.04\;$ms & $25\;$\% & $0.074\;$\% & $1.33\;$\% & & 0.91\\[0.5ex]
       2.0-2.0$\;\sunmass$ & $10\;$Mpc & & 18.4 & 23.9 & 14.1 & 33.3 & & $0.008\;$rad & $0.14\;$ms & $14\;$\% & $0.009\;$\% & $0.14\;$\% & & 0.80\\
       2.0-2.0$\;\sunmass$ & $20\;$Mpc & &  9.2 & 11.9 &  7.1 & 16.7 & & $0.017\;$rad & $0.28\;$ms & $18\;$\% & $0.014\;$\% & $0.23\;$\% & & 0.73\\
       2.0-2.0$\;\sunmass$ & $30\;$Mpc & &  6.1 &  8.0 &  4.7 & 11.1 & & $0.026\;$rad & $0.42\;$ms & $21\;$\% & $0.021\;$\% & $0.37\;$\% & & 0.78\\
    \end{tabular}
  \end{ruledtabular}
  \leftline{\footnotesize $^\mathrm{a}$ spherical standard deviation \cite{MardiaJuppChapter92}}
\end{table*}
One example is the inclination angle~$\iota$, which still 
has a `well-behaving' posterior distribution 
at 10~Mpc distance (see FIG.~\ref{fig:bivariatePosteriors}).
For a weaker signal originating from 30~Mpc distance, 
the distribution then turns bimodal
(FIG.~\ref{fig:ambiguities}). 
The `orientation' of the inclination angle is not clear any more, 
the result being two roughly equally likely 
`mirror image' solutions with 
$\prob(\iota<\frac{\pi}{2}) \approx \frac{1}{2} \approx \prob(\iota>\frac{\pi}{2})$. 
Note that the two solutions $\iota$ and $\pi-\iota$ correspond to opposite orbital directions
(clockwise/counterclockwise), as seen from Earth, 
which might be of minor interest anyway.

The sky location's posterior also exhibits multiple modes 
for this weaker signal (FIG.~\ref{fig:ambiguities}). 
This illustrates some potential pitfalls of Maximum-Likelihood (ML) or 
Maximum-a-Posteriori (MAP) methods; 
these would advise picking the highest of the several modes, 
which might just be the narrowest one, but not necessarily the most likely.
If one then proceeded by extrapolating the curvature at that mode
and deriving error bounds from the Fisher Information matrix,
the resulting estimates might not only be far off, but also associated with overestimated accuracies.

We also tried MCMC runs with a modified prior setting; we extended the prior 
for the coalescence time~$t_c$ from its original range of 
$\pm 5 \mbox{ms}$ around the true value 
to $\pm 27 \mbox{ms}$, allowing for an additional margin of $22 \mbox{ms}$, 
which is the time it takes a gravitational wave to travel from 
Earth's surface to its center. 
This setting reflects the case where the inspiral detection pipeline 
received triggers from less than three interferometer sites, 
so the signal's arrival time at the geocenter could not be estimated 
to greater accuracy in advance.
The MCMC algorithm is still able to find the mode in the 
enlarged time parameter range, but takes more iterations to converge.

\begin{figure}[t]
  \includegraphics[width=8cm]{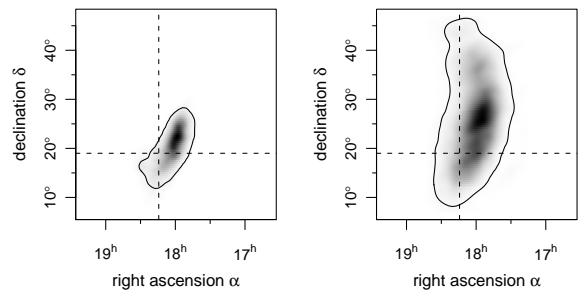}
  \caption{\label{fig:extremeSNR}
           Even if the SNR at one of the interferometers is almost zero, 
           it still contributes to estimates' accuracies---the posterior 
           is much narrower if its data is included (left plot)
           than if it is omitted (right plot).
           (Dashed lines indicate the true values.)}
\end{figure}

One scenario in which such an approach would be necessary is
when the SNR for one of the interferometers is almost zero. 
In such a case 
the data from the interferometer under consideration
also would not (directly) contribute to 
the estimation of phase- and frequency-related parameters, 
but would still carry information about amplitude-related parameters---by implicitly 
`ruling out' those parameter combinations that \emph{would} have resulted in a response at that 
interferometer.
FIG.~\ref{fig:extremeSNR} shows the sky location posteriors for such a signal, 
a \mbox{1.5-2.0~$\sunmass$} inspiral at 10~Mpc distance,
where the SNRs at the three interferometer sites are:
LHO:~9.6, LLO:~13.9, V:~0.18 (network:~16.9).
Including the data from the third interferometer (with almost zero SNR)
into the analysis
still yields a much more accurate estimate of the sky location.
Table~\ref{tab:extremeSNR} compares the resulting parameter accuracies of these two settings.
The posteriors for sky location $(\delta,\alpha)$ and coalescence time~$t_c$, 
which are closely related, are much narrower when the Virgo data is considered in the analysis, 
while estimates of the rather phase- and frequency-related parameters~$m_c$ 
and~$\mu$ do not gain from the additional information.

On the one hand, not only a high (total) SNR is desirable 
but also one that is rather `evenly spread' over different interferometers.
On the other hand even a near-zero SNR at one of the interferometers does not make its measurement useless. 
Inference on different parameters will be affected to different degrees 
by such an unbalanced SNR arrangement.

\begin{table}[h]
  \caption{\label{tab:extremeSNR} Relative accuracies of different parameters 
           (in analogy to Table~\ref{tab:precisions})
           when considering / not considering the Virgo data 
           %for which the SNR is almost zero here.}
           (where the example situation is such that the SNR is nearly zero).}
  \begin{ruledtabular}
  \begin{tabular}{rccccc}
    Virgo data... & $(\delta,\alpha)$ &    $t_c$   &  $d_L$   &   $m_c$     &  $\mu$     \\ \hline
    ...included  & $0.071\;$rad         & $0.81\;$ms & $21\;$\% & $0.023\;$\% & $0.33\;$\% \\
    ...excluded  & $0.150\;$rad         & $2.38\;$ms & $23\;$\% & $0.022\;$\% & $0.31\;$\% 
  \end{tabular}
  \end{ruledtabular}
\end{table}

\section{\label{sec:Discuss}Discussion}
We have developed a new MCMC program for estimating the nine parameters
associated with an inspiral of compact binary objects from the data
coming from a network of gravitational wave interferometers. The
determination of the sky location of the source is an important
consequence of the procedure. Numerous new features have been
implemented in this binary inspiral MCMC\@. The MCMC uses waveform
approximations
%that are 2.0~PN in amplitude and 2.5~PN in phase 
that are 2.5~PN in phase and 2.0~PN in amplitude 
\cite{Blanchet2001}
%(and our code can be easily extended to signal
%waveforms that are 3.5~PN in phase and 2.5~PN in amplitude
%\cite{BlanchetEtAl2002,ArunEtAl2004}).
(and by the time of final submission of this paper 
a version using waveforms that are 3.5~PN in phase 
and 2.5~PN in amplitude 
%\cite{BlanchetEtAl2002,ArunEtAl2004} 
\cite{BlanchetEtAl2002,BlanchetEtAl2005a,ArunEtAl2004,ArunEtAl2005,BlanchetEtAl2004}
was running as well).
The data from multiple interferometers (two or more) are
coherently analyzed in order to produce posterior probability 
distributions for all nine parameters. 

Advanced MCMC techniques were implemented in our program in order to
maximize the efficiency of converging to the correct parameter values in
the large, \mbox{9-}dimensional, parameter space. The initial parameter values
for the sampler 
%are 
were determined using importance resampling \cite{BDA}. 
We recently extended 
(though not with the results presented in this paper)
the parallel tempering algorithm
to an Evolutionary MCMC algorithm \cite{LiangWong2001a}. This MCMC
variety implements proposals that are motivated by genetic
algorithms \cite{Goldberg}, and so recombinations of parameter
samples from different MCMC chains are used as proposals
in order to improve convergence and mixing.

Another current related research effort is the application of a version
of this MCMC code to burst waveforms.
This problem is by orders of magnitude computationally less expensive,
due to the much shorter duration of the signals.
But it appears that on the other hand convergence, 
i.e.\ finding the main posterior mode in the parameter space, 
still poses a major problem.
The theoretical background of the various potential burst sources
is rather vague, so realistic waveforms and 
sensible specifications of parameterisations
and priors also need to be identified. 
The results of this study on the MCMC parameter
estimation of burst signals using the coherent analysis of 
multi-interferometer data will be presented in a forthcoming publication.

We are also extending the MCMC techniques used in this study to the
application of data analysis for LISA detection of binary inspiral
signals. While our present program coherently analyzes data from
multiple ground based interferometers, we have found it is a straightforward 
extension of the code so that we can coherently analyze the time
delay interferometry data from LISA. These results are also forthcoming.

Presently LIGO is at its target sensitivity. Virgo is fast approaching
its design sensitivity. Using the LIGO-Virgo network it will be possible
to observe neutron star binary inspirals out to a distance of 35 Mpc
\cite{Sigg2004,AcerneseEtAl2005}.
A detection of such an inspiral could occur at any time \cite{KalogeraEtAl2006}. 
As displayed in this paper, our MCMC routine is capable of coherently analyzing 
the data from the multiple interferometers, 
and then using it to estimate the nine parameters associated with such a signal.

\begin{acknowledgments}
This work was supported by 
The Royal Society of New Zealand Marsden Fund grant \mbox{UOA-204},
National Science Foundation grant \mbox{PHY-0553422},
and the Fulbright Scholar Program.
\end{acknowledgments}

\bibliography{../../literature/literature}

\end{document}